\documentclass[10pt]{IEEEtran}
\usepackage{amsmath,amssymb}
\usepackage{cite}
\usepackage{paralist}

\DeclareMathOperator{\spann}{span}
 \DeclareMathOperator{\tr}{tr}

 \DeclareMathOperator{\swt}{swt}
\DeclareMathOperator{\wt}{wt}

\newcommand{\C}{\mathbb{C}}
\newcommand{\F}{\mathbb{F}}
\newcommand{\nix}[1]{}
\newcommand{\sdual}{{\perp_s}}

\newcommand{\hdual}{{\perp_h}}

\newcommand{\scal}[2]{\langle #1\mid #2\rangle_s}
\newtheorem{theorem}{\textbf{Theorem}}
\newtheorem{corollary}[theorem]{\textbf{Corollary}}
\newtheorem{lemma}[theorem]{\textbf{Lemma}}

\newtheorem{remark}{\textbf{Remark}}

\begin{document}
\title{Subsystem Code Constructions}
\author{
\authorblockN{Salah A. Aly and Andreas Klappenecker\\}
\authorblockA{Department  of Computer Science \\
Texas A\&M University,
College Station, TX 77843-3112, USA \\
Email: \{salah, klappi\}@cs.tamu.edu} } \maketitle

\begin{abstract}
Subsystem codes are the most versatile class of quantum
error-correcting codes known to date that combine the best features of
all known passive and active error-control schemes.  The subsystem
code is a subspace of the quantum state space that is decomposed into a
tensor product of two vector spaces: the subsystem and the
co-subsystem.  A generic method to derive subsystem codes from
existing subsystem codes is given that allows one to trade the
dimensions of subsystem and co-subsystem while maintaining or
improving the minimum distance. As a consequence, it is shown that all
pure MDS subsystem codes are derived from MDS stabilizer codes.  The
existence of numerous families of MDS subsystem codes is established.
Propagation rules are derived that allow one to obtain longer and
shorter subsystem codes from given subsystem codes. Furthermore,
propagation rules are derived that allow one to construct a new
subsystem code by combining two given subsystem codes.
\end{abstract}
\section{Introduction}
Subsystem codes are a relatively new construction of quantum codes
that combine the features of decoherence free
subspaces~\cite{lidar98}, noiseless subsystems~\cite{zanardi97}, and
quantum error-correcting codes~\cite{calderbank98,gottesman96}. Such
codes promise to offer appealing features, such as simplified syndrome
calculation and a wide variety of easily implementable fault-tolerant
operations, see~\cite{aliferis06,aly06c,bacon06,kribs05}.

An $((n,K,R,d))_q$ subsystem code is a $KR$-dimensional subspace $Q$
of $\C^{q^n}$ that is decomposed into a tensor product $Q=A\otimes B$
of a $K$-dimensional vector space $A$ and an $R$-dimensional vector
space $B$ such that all errors of weight less than~$d$ can be detected
by~$A$. The vector spaces $A$ and $B$ are respectively called the
subsystem $A$ and the co-subsystem $B$. For some background on
subsystem codes, see for instance~\cite{pre0608,poulin05,aly06c}.

A special feature of subsystem codes is that any classical additive
code $C$ can be used to construct a subsystem code. One should
contrast this with stabilizer codes, where the classical codes are
required to satisfy a self-orthogonality condition.

We assume that the reader is familiar with the relation between
classical and quantum stabilizer codes,
see~\cite{calderbank98,rains99}. In \cite{aly06c,pre0608}, the
authors gave an introduction to subsystem codes, established upper and
lower bounds on subsystem code parameters, and provided two methods
for constructing subsystem codes. The main results on this paper are
as follows:
\begin{compactenum}[i)]
\item If $q$ is a power of a prime $p$, then we show that a subsystem
code with parameters $((n,K/p,pR,\geq d))_q$ can be obtained from a
subsystem code with parameters $((n,K,R,d))_q$. Furthermore, we show
that the existence of a pure $((n,K,R,d))_q$ subsystem code implies
the existence of a pure $((n,pK,R/p,d))_q$ code.
\item We show that all pure MDS subsystem codes are derived from MDS
stabilizer codes. We establish here for the first time the existence
of numerous families of MDS subsystem codes.
\item We derive two propagation rules that yield new subsystem
codes by extending or shortening the length existing codes.
\item We derive two propagation rules that yield a new subsystem code
by combining two subsystem codes.
\end{compactenum}

\section{Subsystem Code Constructions}
First we recall the following fact that is key to most constructions of
subsystem codes (see below for notations):
\begin{theorem}\label{th:oqecfq}
Let $C$ be a classical additive subcode of\/ $\F_q^{2n}$ such that
$C\neq \{0\}$ and let $D$ denote its subcode $D=C\cap C^\sdual$. If
$x=|C|$ and $y=|D|$, then there exists a subsystem code $Q= A\otimes
B$ such that
\begin{compactenum}[i)]
\item $\dim A = q^n/(xy)^{1/2}$,
\item $\dim B = (x/y)^{1/2}$.
\end{compactenum}
The minimum distance of subsystem $A$ is given by
\begin{compactenum}[(a)]
\item $d=\swt((C+C^\sdual)-C)=\swt(D^\sdual-C)$ if $D^\sdual\neq C$;
\item $d=\swt(D^\sdual)$ if $D^\sdual=C$.
\end{compactenum}
Thus, the subsystem $A$ can detect all errors in $E$ of weight less
than $d$, and can correct all errors in $E$ of weight $\le \lfloor
(d-1)/2\rfloor$.
\end{theorem}
\begin{IEEEproof}
See~\cite[Theorem~5]{pre0608}.
\end{IEEEproof}

A subsystem code that is derived with the help of the previous theorem
is called a Clifford subsystem code. We will assume throughout this
paper that all subsystem codes are Clifford subsystem codes. In
particular, this means that the existence of an $((n,K,R,d))_q$
subsystem code implies the existence of an additive code $C\le
\F_q^{2n}$ with subcode $D=C\cap C^\sdual$ such that $|C|=q^nR/K$,
$|D|=q^n/(KR)$, and $d=\swt(D^\sdual - C)$.

A subsystem code derived from an additive classical code $C$ is called
pure to $d'$ if there is no element of symplectic weight less than
$d'$ in $C$. A subsystem code is called pure if it is pure to the
minimum distance $d$. We require that an $((n,1,R,d))_q$ subsystem
code must be pure.

We also use the bracket notation $[[n,k,r,d]]_q$ to write the
parameters of an $((n,q^k,q^r,d))_q$ subsystem code in simpler form.
Some authors say that an $[[n,k,r,d]]_q$ subsystem code has $r$
gauge qudits, but this terminology is slightly confusing, as the
co-subsystem typically does not correspond to a state space of $r$
qudits except perhaps in trivial cases. We will avoid this
misleading terminology. An $((n,K,1,d))_q$ subsystem code is also an
$((n,K,d))_q$ stabilizer code and vice versa.

{\em Notation.} Let $q$ be a power of a prime integer $p$. We denote
by $\F_q$ the finite field with $q$ elements. We use the notation
$(x|y)=(x_1,\dots,x_n|y_1,\dots,y_n)$ to denote the concatenation of
two vectors $x$ and $y$ in $\F_q^n$. The symplectic weight of
$(x|y)\in \F_q^{2n}$ is defined as $$\swt(x|y)=\{(x_i,y_i)\neq
(0,0)\,|\, 1\le i\le n\}.$$
We define $\swt(X)=\min\{\swt(x)\,|\, x\in X, x\neq 0\}$ for any nonempty
subset $X\neq \{0\}$ of $\F_q^{2n}$.

The trace-symplectic product of two
vectors $u=(a|b)$ and $v=(a'|b')$ in $\F_q^{2n}$ is defined as
$$\langle u|v \rangle_s = \tr_{q/p}(a'\cdot b-a\cdot b'),$$ where
$x\cdot y$ denotes the dot product and $\tr_{q/p}$ denotes the trace
from $\F_q$ to the subfield $\F_p$.  The trace-symplectic dual of a
code $C\subseteq \F_q^{2n}$ is defined as $$C^\sdual=\{ v\in
\F_q^{2n}\mid \langle v|w \rangle_s =0 \mbox{ for all } w\in C\}.$$
We define the Euclidean inner product $\langle x|y\rangle
=\sum_{i=1}^nx_iy_i$ and the Euclidean dual of $C\subseteq \F_{q}^n$
as $$C^\perp = \{x\in \F_{q}^n\mid \langle x|y \rangle=0 \mbox{ for
all } y\in C \}.$$ We also define the Hermitian inner product for
vectors $x,y$ in $\F_{q^2}^n$ as $\langle x|y\rangle_h
=\sum_{i=1}^nx_i^qy_i$ and the Hermitian dual of $C\subseteq
\F_{q^2}^n$ as
$$C^\hdual= \{x\in \F_{q^2}^n\mid \langle x|y \rangle_h=0 \mbox{ for all } y\in
C \}.$$

\section{Trading Dimensions of subsystem and co-subsystem codes}\label{sec:dimensions}
In this section we show how one can trade the dimensions of subsystem and
co-subsystem to obtain new codes from a given subsystem or stabilizer code.
The results are obtained by exploiting the symplectic geometry of the space. A
remarkable consequence is that nearly any stabilizer code yields a series of
subsystem codes.

Our first result shows that one can decrease the dimension of the
subsystem and increase at the same time the dimension of the
co-subsystem while keeping or increasing the minimum distance of the
subsystem code.

\begin{theorem}\label{th:shrinkK}
Let $q$ be a power of a prime~$p$. If there exists an $((n,K,R,d))_q$
subsystem code with $K>p$ that is pure to $d'$, then there exists an
$((n,K/p,pR,\geq d))_q$ subsystem code that is pure to $\min\{d,d'\}$.
If a pure $((n,p,R,d))_q$ subsystem code exists, then there exists a
$((n,1,pR,d))_q$ subsystem code.
\end{theorem}
\begin{IEEEproof}
By definition, an $((n,K,R,d))_q$ Clifford subsystem code is
associated with a classical additive code $C \subseteq \F_q^{2n}$ and
its subcode $D=C\cap C^\sdual$ such that $x=|C|$, $y=|D|$,
$K=q^n/(xy)^{1/2}$, $R=(x/y)^{1/2}$, and $d=\swt(D^\sdual - C)$ if
$C\neq D^\sdual$, otherwise $d=\swt(D^\sdual)$ if $D^\sdual=C$.

We have $q=p^m$ for some positive integer $m$. Since $K$ and $R$ are
positive integers, we have $x=p^{s+2r}$ and $y=p^s$ for some integers
$r\ge 1$, and $s\ge 0$. There exists an $\F_p$-basis of $C$
of the form
$$ C = \spann_{\F_p}\{z_1,\dots,z_s,x_{s+1},z_{s+1},\dots,
x_{s+r},z_{s+r}\}$$ that can be extended to a symplectic basis
$\{x_1,z_1,\dots,x_{nm},z_{nm}\}$ of $\F_q^{2n}$, that is,
$\scal{x_k}{x_\ell}=0$, $\scal{z_k}{z_\ell}=0$,
$\scal{x_k}{z_\ell}=\delta_{k,\ell}$ for all $1\le k,\ell \le nm$,
see~\cite[Theorem 8.10.1]{cohn05}.

Define an additive code $$C_m =
\spann_{\F_p}\{z_1,\dots,z_s,x_{s+1},z_{s+1},\dots,
x_{s+r+1},z_{s+r+1}\}.$$ It follows that
$$C^\sdual_m=\spann_{\F_p}\{z_1,\dots,z_s,x_{s+r+2},z_{s+r+2}, \dots,
x_{nm},z_{nm}\}$$ and
$$D=C_m\cap C_m^\sdual =
\spann_{\F_p}\{z_1,\dots,z_s\}.$$
By definition, the code $C$ is a
subset of $C_m$.

The subsystem code defined by $C_m$ has the parameters
$(n,K_m,R_m,d_m)$, where $K_m=q^n/(p^{s+2r+2}p^s)^{1/2}=K/p$ and
$R_m=(p^{s+2r+2}/p^s)^{1/2}=pR$. For the claims concerning
minimum distance and purity, we distinguish two cases:
\begin{compactenum}[(a)]
\item If $C_m\neq D^\sdual$, then $K>p$ and $d_m=\swt(D^\sdual -
C_m)\ge \swt(D^\sdual-C)=d$. Since by hypothesis $\swt(D^\sdual-C)=d$
and $\swt(C)\ge d'$, and $D\subseteq C\subset C_m\subseteq D^\sdual$
by construction, we have $\swt(C_m)\ge \min\{ d,d'\}$; thus, the
subsystem code is pure to $\min\{d,d'\}$.

\item If $C_m=D^\sdual$, then $K_m=1=K/p$, that is, $K=p$;  it follows from
the assumed purity that $d=\swt(D^\sdual-C)=\swt(D^\sdual)=d_m$.
\end{compactenum}
This proves the claim.
\end{IEEEproof}

For $\F_q$-linear subsystem codes there exists a variation of the
previous theorem which asserts that one can construct the resulting
subsystem code such that it is again $\F_q$-linear.

\begin{theorem}\label{th:FqshrinkK}
Let $q$ be a power of a prime~$p$. If there exists an $\F_q$-linear
$[[n,k,r,d]]_q$ subsystem code with $k>1$ that is pure to $d'$, then
there exists an $\F_q$-linear $[[n,k-1,r+1,\geq d]]_q$ subsystem code
that is pure to $\min\{d,d'\}$.  If a pure $\F_q$-linear
$[[n,1,r,d]]_q$ subsystem code exists, then there exists an
$\F_q$-linear $[[n,0,r+1,d]]_q$ subsystem code.
\end{theorem}
\begin{IEEEproof}
The proof is analogous to the proof of the previous theorem, except
that $\F_q$-bases are used instead of $\F_p$-bases.
\end{IEEEproof}

There exists a partial converse of Theorem~\ref{th:shrinkK}, namely if
the subsystem code is pure, then it is possible to increase the
dimension of the subsystem and decrease the dimension of the
co-subsystem while maintaining the same minimum distance.

\begin{theorem}\label{th:shrinkR}
Let $q$ be a power of a prime $p$. If there exists a pure
$((n,K,R,d))_q$ subsystem code with $R>1$, then there exists a pure
$((n,pK,R/p,d))_q$ subsystem code.
\end{theorem}
\begin{IEEEproof}
Suppose that the $((n,K,R,d))_q$ Clifford subsystem code is associated
with a classical additive code
$$ C_m = \spann_{\F_p}\{z_1,\dots,z_s,x_{s+1},z_{s+1},\dots,
x_{s+r+1},z_{s+r+1}\}.$$ Let $D=C_m\cap C_m^\sdual$. We have
$x=|C_m|=p^{s+2r+2}$, $y=|D|=p^s$, hence $K=q^n/p^{r+s}$ and
$R=p^{r+1}$. Furthermore, $d=\swt(D^\sdual)$.

The code $$C=\spann_{\F_p}\{z_1,\dots,z_s,x_{s+1},z_{s+1},\dots,
x_{s+r},z_{s+r}\}$$ has the subcode $D=C\cap C^\sdual$. Since
$|C|=|C_m|/p^2$, the parameters of the Clifford subsystem code
associated with $C$ are $((n,pK,R/p,d'))_q$. Since $C\subset C_m$, the
minimum distance $d'$ satisfies $$d'=\swt(D^\sdual-C)\le \swt(D^\sdual
- C_m)=\swt(D^\sdual)=d.$$  On the other hand, $d'=\swt(D^\sdual-C)\ge
\swt(D^\sdual)=d$, whence $d=d'$. Furthermore, the resulting code is pure
since $d=\swt(D^\sdual)=\swt(D^\sdual-C)$.
\end{IEEEproof}

Replacing $\F_p$-bases by $\F_q$-bases in the proof of the
previous theorem yields the following variation of the previous
theorem for $\F_q$-linear subsystem codes.
\begin{theorem}\label{th:FqshrinkR}
Let $q$ be a power of a prime $p$. If there exists a pure
$\F_q$-linear $[[n,k,r,d]]_q$ subsystem code with $r>0$, then there
exists a pure $\F_q$-linear $[[n,k+1,r-1,d]]_q$ subsystem code.
\end{theorem}

The purity hypothesis in Theorems~\ref{th:shrinkR}
and~\ref{th:FqshrinkR} is essential, as the next remark shows.

\begin{remark}
The Bacon-Shor code is an impure $[[9,1,4,3]]_2$ subsystem
code. However, there does not exist any $[[9,5,3]]_2$ stabilizer code.
Thus, in general one cannot omit the purity assumption from
Theorems~\ref{th:shrinkR} and~\ref{th:FqshrinkR}.
\end{remark}

An $[[n,k,d]]_q$ stabilizer code can also be regarded as an
$[[n,k,0,d]]_q$ subsystem code. We record this important special case
of the previous theorems in the next corollary.

\goodbreak
\begin{corollary}\label{cor:generic}
If there exists an ($\F_q$-linear) $[[n,k,d]]_q$ stabilizer code that
is pure to $d'$, then there exists for all $r$ in the range $0\le r<k$
an ($\F_q$-linear) $[[n,k-r,r,\ge d]]_q$ subsystem code that is pure
to $\min\{d,d'\}$ .  If a pure ($\F_q$-linear) $[[n,k,r,d]]_q$
subsystem code exists, then a pure ($\F_q$-linear) $[[n,k+r,d]]_q$
stabilizer code exists.
\end{corollary}

\section{MDS Subsystem Codes}
Recall that an $[[n,k,r,d]]_q$ subsystem code derived from an
$\F_q$-linear classical code $C\le \F_q^{2n}$ satisfies the Singleton
bound $k+r\le n-2d+2$, see~\cite[Theorem~3.6]{pre0703}. A subsystem
code attaining the Singleton bound with equality is called an MDS
subsystem code.

An important consequence of the previous theorems is the following
simple observation which yields an easy construction of subsystem codes
that are optimal among the $\F_q$-linear Clifford subsystem codes.

\begin{theorem}\label{th:pureMDS}
If there exists an $\F_q$-linear $[[n,k,d]]_q$ MDS stabilizer code,
then there exists a pure $\F_q$-linear $[[n,k-r,r,d]]_q$ MDS subsystem
code for all $r$ in the range $0\le r\le k$.
\end{theorem}
\begin{IEEEproof}
An MDS stabilizer code must be pure, see~\cite[Theorem~2]{rains99}
or \cite[Corollary 60]{ketkar06}. By Corollary~\ref{cor:generic}, a
pure $\F_q$-linear $[[n,k,d]]_q$ stabilizer code implies the
existence of an $\F_q$-linear $[[n,k-r,r, d_r\ge d]]_q$ subsystem
code that is pure to~$d$ for any $r$ in the range $0\le r\le k$.
Since the stabilizer code is MDS, we have $k=n-2d+2$. By the
Singleton bound, the parameters of the resulting $\F_q$-linear
$[[n,n-2d+2-r,r,d_r]]_q$ subsystem codes must satisfy
$(n-2d+2-r)+r\le n-2d_r+2$, which shows that the minimum distance
$d_r=d$, as claimed.
\end{IEEEproof}

\begin{remark}
We conjecture that $\F_q$-linear MDS subsystem codes are actually
optimal among all subsystem codes, but a proof that the Singleton
bound holds for general subsystem codes remains elusive.
\end{remark}
\smallskip

In the next lemma, we give a few examples of MDS subsystem codes
that can be obtained from Theorem~\ref{th:pureMDS}. These are the
first families of MDS subsystem codes (though sporadic examples of
MDS subsystem codes have been established before, see
e.g.~\cite{aly06c,bacon06}).
\begin{lemma}
\begin{enumerate}[i)]
\item An $\F_q$-linear pure $[[n,n-2d+2-r,r,d]]_q$ MDS subsystem code exists
for all $n$, $d$, and $r$ such that $3\le n\le q$, $1\le d\le n/2+1$,
and\/ $0\le r\le n-2d+1$.
\item An $\F_q$-linear pure $[[(\nu+1)q,(\nu+1)q-2\nu-2-r,r,\nu+2]]_q$ MDS subsystem code exists for all $\nu$ and $r$ such that $0\le \nu\le q-2$ and
$0\le r\le (\nu+1)q-2\nu-3$.
\item An $\F_q$-linear pure $[[q - 1, q-1 -2\delta -r,
r,\delta + 1]]_q$ MDS subsystem code exists for all
$\delta$ and $r$ such that $0 \leq \delta < (q -1)/2$ and $0\leq r \le q - 2\delta - 1$.
\item An $\F_q$-linear pure $[[q, q -
2\delta - 2-r',r', \delta + 2]]_q$ MDS subsystem code exists for all
$0 \leq \delta < (q -1)/2$ and $0\leq r' <q - 2\delta - 2$.
\item An $\F_q$-linear pure $[[q^2 - 1, q^2 - 2\delta - 1-r,r, \delta +
1]]_q$ MDS subsystem code exists for all $\delta$ and $r$ in the range
$0 \leq \delta < q-1$ and $0\leq r< q^2 - 2\delta - 1$.
\item An $\F_q$-linear pure $[[q^2, q^2 - 2\delta -
2-r',r', \delta + 2]]_q$ MDS subsystem code exists for all $\delta$ and $r'$ in the range
$0 \leq \delta < q-1$ and $0\leq r' <q^2 - 2\delta - 2$.
\end{enumerate}
\end{lemma}
\begin{IEEEproof}
\begin{inparaenum}[i)]
\item By \cite[Theorem~14]{grassl04}, there exist $\F_q$-linear
$[[n,n-2d+2,d]]_q$ stabilizer codes for all $n$ and $d$ such that
$3\le n\le q$ and $1\le d\le n/2+1$. The claim follows from Theorem~\ref{th:pureMDS}. \\
\item By \cite[Theorem~5]{klappenecker050}, there exist a
$[[(\nu+1)q,(\nu+1)q-2\nu-2,\nu+2]]_q$ stabilizer code. In this case,
the code is derived from an $\F_{q^2}$-linear code $X$ of length $n$
over $\F_{q^2}$ such that $X\subseteq X^\hdual$. The claim follows
from Lemma~\ref{l:hermitian-linear} and Theorem~\ref{th:pureMDS}.\\
\item$\!\!$,\item There exist $\F_q$-linear stabilizer codes with
parameters $[[q - 1, q - 2\delta - 1,\delta + 1]]_q$ and $[[q, q -
2\delta - 2, \delta + 2]]_q$ for $0 \leq \delta < (q -1)/2$, \
see~\cite[Theorem~9]{grassl04}. Theorem~\ref{th:pureMDS} yields the claim. \\
\item$\!\!$,\item There exist $\F_q$-linear stabilizer codes with
parameters $[[q^2 - 1, q^2 - 2\delta - 1, \delta + 1]]_q$ and $[[q^2,
q^2 - 2\delta - 2, \delta + 2]]_q$.  for $0 \leq \delta < q-1$
by~\cite[Theorem~10]{grassl04}. The claim follows from
Theorem~\ref{th:pureMDS}.
\end{inparaenum}
\end{IEEEproof}
The existence of the codes in i) are merely established by a
non-constructive Gilbert-Varshamov type counting argument.  However,
the result is interesting, as it asserts that there exist for example
$[[6,1,1,3]]_q$ subsystem codes for all prime powers $q\ge 7$,
$[[7,1,2,3]]_q$ subsystem codes for all prime powers $q\ge 7$, and
other short subsystem codes that one should compare with a
$[[5,1,3]]_q$ stabilizer code. If the syndrome calculation is simpler,
then such subsystem codes could be of practical value.

The subsystem codes given in ii)-vi) of the previous lemma are
constructively established. The subsystem codes in ii) are derived
from Reed-Muller codes, and in iii)-vi) from Reed-Solomon codes.
There exists an overlap between the parameters given in ii) and in
iv), but we list here both, since each code construction has its own
merits.
\begin{remark}
By Theorem~\ref{th:FqshrinkR}, pure MDS subsystem codes can always
be derived from MDS stabilizer codes, see Table~\ref{table:optimalMDS}. Therefore,
one can derive in fact all possible parameter sets of pure MDS
subsystem codes with the help of Theorem~\ref{th:pureMDS}.
\end{remark}
\begin{remark}
In the case of stabilizer codes, all MDS codes must be pure. For
subsystem codes this is not true, as the $[[9,1,4,3]]_2$ subsystem
code shows. Finding such impure $\F_q$-linear $[[n,k,r,d]]_q$ MDS
subsystem codes with $k+r=n-2d+2$ is a particularly interesting
challenge.
\end{remark}

\begin{table}[t]
\caption{Optimal pure subsystem codes} \label{table:optimalMDS}
\begin{center}
\begin{tabular}{|c|c|c|}
\hline
\text{Subsystem Codes} &  \text{Parent}  \\
 &  \text{Code (RS Code)}  \\
\hline $[[8  ,1  ,5  ,2  ]]_3$ & $[8  ,6  ,3  ]_{3^2}$ \\{}

$[[8  ,4  ,2  ,2  ]]_3$&$[8  ,3  ,6  ]_{3^2}$\\{} $[[8  ,5  ,1  ,2
]]_3$&$[8 ,2 ,7 ]_{3^2}$\\{}

$[[9  ,1 ,4 ,3 ]]_3$&$[9  ,6 ,4 ]_{3^2}^{\dag}, \delta=3$\\
$[[9  ,4 ,1 ,3 ]]_3$&$[9  ,3 ,7 ]_{3^2}^{\dag}, \delta=6$ \\
 \hline $[[15,1,10,3]]_4$ &
$[15 ,12 ,4 ]_{4^2}$  \\{}
$[[15,9,2,3]]_4$&$[15,4,12]_{4^2}$\\{} $[[15,10,1,3]]_4$&$[15,3,13]_{4^2}$\\
$[[16 ,1 ,9 ,4 ]]_4$& $[16 ,12,5 ]_{4^2}^{\dag},\delta= 4 $\\

  \hline $[[24,1,17,4]]_5$
&$[24,20,5]_{5^2}$
\\{}
$ [[24,16,2,4]]_5$ &$[24,5,20]_{5^2}$\\{} $[[24,17 ,1,4 ]]_5
$&$[24,4,21]_{5^2}$\\{}
 $[[24,19,1,3]]_5$ &$[24,3,22]_{5^2}$\\{}
$[[24 ,21 ,1  ,2  ]]_5$ & $ [24 ,2  ,23 ]_{5^2}$ \\{} $[[23 ,1 ,18,3
]]_5$&$[23 ,20,4 ]_{5^2}^{*}, \delta=5$\\{}
$[[23 ,16,3 ,3 ]]_5$&$[23 ,5 ,19]_{5^2}^{*}, \delta=20$\\
 \hline
 $[[48 ,1  ,37 ,6  ]]_7$  &$[48 ,42 ,7 ]_{7^2}$\\
 \hline
\end{tabular}
\\
* Punctured code\\
$\dag$ Extended code
\end{center}
 \end{table}

Recall that a pure subsystem code is called perfect if and only if it
attains the Hamming bound with equality. We conclude this section with
the following consequence of Theorem~\ref{th:pureMDS}:
\begin{corollary}
If there exists an $\F_q$-linear pure $[[n,k,d]]_q$ stabilizer code
that is perfect, then there exists a pure $\F_q$-linear
$[[n,k-r,r,d]]_q$ perfect subsystem code for all $r$ in the range
$0\leq r \leq k$.
\end{corollary}

\section{Extending and Shortening Subsystem Codes}
In Section~\ref{sec:dimensions}, we showed how one can derive new
subsystem codes from known ones by modifying the dimension of the
subsystem and co-subsystem. In this section, we derive new subsystem
codes from known ones by extending and shortening the length of the
code.

\begin{theorem}\label{lemma_n+1k}
If there exists an $((n,K,R,d))_q$ Clifford subsystem  code
with $K>1$, then there exists an $((n+1, K, R, \ge d))_q$
subsystem code that is pure to~1.
\end{theorem}
\begin{IEEEproof}
We first note that for any additive subcode $X\le \F_q^{2n}$, we can
define an additive code
$X'\le \F_q^{2n+2}$ by
$$X'=\{ (a\alpha|b0)\,|\, (a|b)\in X, \alpha\in
\F_q\}.$$ We have $|X'|=q|X|$. Furthermore, if $(c|d)\in X^\sdual$,
then $(c\alpha|d0)$ is contained in $(X')^\sdual$ for all $\alpha$ in
$\F_q$, whence $(X^\sdual)'\subseteq (X')^\sdual$. By comparing
cardinalities we find that equality must hold; in other words, we have
$$(X^\sdual)'= (X')^\sdual.$$

By Theorem~\ref{th:oqecfq}, there are two additive codes $C$ and $D$
associated with an $((n,K,R,d))_q$ Clifford subsystem code such that
$$|C|=q^nR/K$$ and $$|D|=|C\cap C^\sdual| = q^n/(KR).$$ We can derive from the
code $C$ two new additive codes of length $2n+2$ over $\F_q$, namely $C'$ and
$D'=C'\cap (C')^\sdual$. The codes $C'$ and $D'$ determine a
$((n+1,K',R',d'))_q$ Clifford subsystem code. Since
\begin{eqnarray*}
D'&=&C'\cap (C')^\sdual = C'\cap (C^\sdual)' \\&=&(C\cap C^\sdual)',
\end{eqnarray*}
 we have
$|D'|=q|D|$. Furthermore, we have $|C'|=q|C|$. It follows from
Theorem~\ref{th:oqecfq} that
\begin{compactenum}[(i)]
\item $K'= q^{n+1}/\sqrt{|C'||D'|}=q^n/\sqrt{|C||D|}=K$,
\item $R'=(|C'|/|D'|)^{1/2} = (|C|/|D|)^{1/2} = R$,
\item $d'= \swt( (D')^\sdual \setminus C')\ge \swt( (D^\sdual\setminus C)')=d$.
\end{compactenum}
Since $C'$ contains a vector $(\mathbf{0}\alpha|\mathbf{0}0)$ of
weight $1$, the resulting subsystem code is pure to~1.
\end{IEEEproof}


\begin{corollary}
If there exists an $[[n,k,r,d]]_q$ subsystem  code with $k>0$ and $0\leq r <k$,
then there exists an $[[n+1, k, r, \ge d]]_q$ subsystem code that is pure to~1.
\end{corollary}
\medskip

We can also shorten the length of a subsystem code in a simple way
as shown in the following Theorem.

\begin{theorem}\label{lem:n-1k+1rule}
If a pure $((n,K,R,d))_q$ subsystem code exists, then there exists
a pure $((n-1,qK,R,d-1))_q$ subsystem code.
\end{theorem}
\begin{IEEEproof}
By \cite[Lemma~10]{aly06c}, the existence of a pure Clifford subsystem
code with parameters $((n,K,R,d))_q$ implies the existence of a pure
$((n,KR,d))_q$ stabilizer code. It follows from~\cite[Lemma
70]{ketkar06} that there exist a pure $((n-1,qKR,d-1))_q$ stabilizer
code, which can be regarded as a pure $((n-1,qKR,1,d-1))_q$ subsystem
code. Thus, there exists a pure $((n-1,qK,R,d-1))_q$ subsystem code by
Theorem~\ref{th:shrinkR}, which proves the claim.
\end{IEEEproof}

In bracket notation, the previous theorem states that the existence of
a pure $[[n,k,r,d]]_q$ subsystem code implies the existence of a pure
$[[n-1,k+1,r,d-1]]_q$ subsystem code.

\section{Combining Subsystem Codes}
In this section, we show how one can obtain a new subsystem code by
combining two given subsystem codes in various ways.

\begin{theorem}\label{thm:twocodes_n1k1r1d1n2k2r2d2}
If there exists a pure $[[n_1,k_1,r_1,d_1]]_2$ subsystem code and a
pure $[[n_2,k_2,r_2,d_2]]_2$ subsystem code such that $k_2+r_2\leq n_1$,
then there exist subsystem codes with parameters
$$[[n_1+n_2-k_2-r_2,k_1+r_1-r,r,d]]_2$$
for all $r$ in the range $0\le r< k_1+r_1$, where
the minimum distance $d \geq \min\{d_1,d_1+d_2-k_2-r_2\}$.
\end{theorem}
\begin{IEEEproof}
Since there exist pure $[[n_1,k_1,r_1,d_1]]_2$ and
$[[n_2,k_2,r_2,d_2]]_2$ subsystem codes with $k_2+r_2\leq n_1$, it
follows from Theorem~\ref{th:shrinkR} that there exist stabilizer
codes with the parameters $[[n_1,k_1+r_1,d_1]]_2$ and
$[[n_2,k_2+r_2,d_2]]_2$ such that $k_2+r_2\leq n_1$. Therefore,
there exists an $[[n_1+n_2-k_2-r_2,k_1+r_1,d]]_2$ stabilizer code
with minimum distance $d \geq \min \{d_1,d_1+d_2-k_2-r_2\}$
by~\cite[Theorem 8]{calderbank98}. It follows from
Theorem~\ref{th:shrinkK} that there exists
$[[n_1+n_2-k_2-r_2,k_1+r_1-r,r,\geq d]]_2$ subsystem codes for all
$r$ in the range $0\le r< k_1+r_1$.
\end{IEEEproof}

\begin{theorem}\label{lem:twocodes_nk1r1s1k2r2d2} Let $Q_1$ and $Q_2$ be two
pure subsystem codes with parameters $[[n,k_1,r_1,d_1]]_q$ and
$[[n,k_2,r_2,d_2]]_q$, respectively. If $Q_2\subseteq Q_1$, then
there exists pure subsystem codes with parameters
$$[[2n,k_1+k_2+r_1+r_2-r,r,d]]_q$$
for all $r$ in the range $0\le r\le k_1+k_2+r_1+r_2$, where
the minimum distance $d \geq \min \{d_1,2d_2\}$.
\end{theorem}
\begin{IEEEproof}
By assumption, there exists a pure $[[n,k_i,r_i,d_i]]_q$ subsystem
code, which implies the existence of a pure $[[n,k_i+r_i,d_i]]_q$
stabilizer code by Theorem~\ref{th:shrinkR}, where $i\in\{1,2\}$.
By~\cite[Lemma 74]{ketkar06}, there exists a pure stabilizer code with
parameters $[[2n,k_1+k_2+r_1+r_2,d]]_q$ such that $d \geq \min
\{2d_2,d_1\}$. By Theorem~\ref{th:shrinkK}, there exist a pure
subsystem code with parameters $[[2n,k_1+k_2+r_1+r_2-r,r,d]]_q$ for
all $r$ in the range $0\le r\le k_1+k_2+r_1+r_2$, which proves the claim.
\end{IEEEproof}

\section{Conclusions and Open Problems}
Subsystem codes -- or operator quantum error-correcting codes as
some authors prefer to called them -- are among the most versatile
tools in quantum error-correction, since they allow one to combine
the passive error-correction found in decoherence free subspaces and
noiseless subsystems with the active error-control methods of
quantum error-correcting codes. The subclass of Clifford subsystem
codes that was studied in this paper is of particular interest
because of the close connection to classical error-correcting codes.
\nix{ As Theorem~\ref{th:oqecfq} shows, one can derive from each
additive code over $\F_q$ an Clifford subsystem code. This offers
more flexibility than the slightly rigid framework of stabilizer
codes.}

In this paper, we showed that any $\F_q$-linear MDS stabilizer code
yields a series of pure $\F_q$-linear MDS subsystem codes. These
codes are known to be optimal among the $\F_q$-linear Clifford
subsystem codes. We conjecture that the Singleton bound holds in
general for subsystem codes. \nix{There is quite some evidence for
this fact, as pure Clifford subsystem codes and $\F_q$-linear
Clifford subsystem codes are known to obey this bound.}

We have established a number of subsystem code constructions. In
particular, we have shown how one can derive subsystem codes from
stabilizer codes. In combination with the propagation rules that we
have derived, one can easily create tables with the best known
subsystem codes. Further propagation rules and examples of such tables
will be given in an expanded version of this paper that is not limited
by space constraints.

\section{ACKNOWLEDGMENTS} This research was supported by NSF grant
CCF-0622201 and NSF CAREER award CCF-0347310. We thank Daniel Lidar
for providing us with references on DFS and noiseless
subsystems. S.A.A. thanks the organizers of the first international
conference on quantum error correction (held at the USC campus,
December 17-21, 2007) for their hospitality.

\appendix We recall that the Hermitian construction of stabilizer
codes yields $\F_q$-linear stabilizer codes, as can be seen from the
following reformulation of~\cite[Corollary~2]{grassl04}.
\begin{lemma}[\cite{grassl04}]\label{l:hermitian-linear}
If there exists an $\F_{q^2}$-linear code $X\subseteq \F_{q^2}^n$ such
that $X\subseteq X^\hdual$, then there exists an $\F_q$-linear code
$C\subseteq \F_q^{2n}$ such that $C\subseteq C^\sdual$, $|C|=|X|$,
$\swt(C^\sdual - C)=\wt(X^\hdual - X)$ and $\swt(C)=\wt(X)$.
\end{lemma}
\begin{IEEEproof}
Let $\{1,\beta\}$ be a basis of $\F_{q^2}/\F_q$. Then
$\tr_{q^2/q}(\beta)=\beta+\beta^q$ is an element $\beta_0$ of $\F_q$; hence,
$\beta^q=-\beta+\beta_0$. Let $$C=\{ (u|v)\,|\, u,v\in \F_q^n, u+\beta v\in
X\}.$$ It follows from this definition that $|X|=|C|$ and that
$\wt(X)=\swt(C)$. Furthermore, if $u+\beta v$ and $u'+\beta v'$ are elements of
$X$ with $u,v,u',v'$ in $\F_q^n$, then
$$
\begin{array}{lcl}
0&=&(u+\beta v)^q\cdot (u'+\beta v') \\
&=& u\cdot u' + \beta^{q+1} v\cdot
v' + \beta_0 v \cdot u' + \beta (u\cdot v' -v \cdot u').
\end{array}
$$ On the right hand side, all terms but the last are in $\F_q$; hence
we must have $(u\cdot v' -v \cdot u')=0$, which shows that $(u|v)
\,\sdual\, (u'|v')$, whence $C\subseteq C^\sdual$. Expanding
$X^\hdual$ in the basis $\{1\,\beta\}$ yields a code $C'\subseteq
C^\sdual$, and we must have equality by a dimension argument. Since
the basis expansion is isometric, it follows that $\swt(C^\sdual -
C)=\wt(X^\hdual - X).$ The $\F_q$-linearity of $C$ is a direct
consequence of the definition of $C$.
\end{IEEEproof}
\newcommand{\XXstud}{{}}
\newcommand{\XXar}[1]{}
\scriptsize
\bibliographystyle{ieeetr}
%

\end{document}